# Team NeuroPoly: Description of the Pipelines for the MICCAI 2021 MS New Lesions Segmentation Challenge

Uzay Macar[1,2*], Enamundram Naga Karthik[1,2*], Charley Gros[1,2], Andréanne Lemay[1,2], Julien Cohen-Adad[1,2,3]

[1] NeuroPoly Lab, Institute of Biomedical Engineering, Polytechnique Montréal, Montréal, QC, Canada

[2] MILA - Québec AI Institute, Montréal, QC, Canada

[3] Functional Neuroimaging Unit, CRIUGM, Université de Montréal, Montréal, QC, Canada

**Abstract.** This paper gives a detailed description of the pipelines used for the challenge including the data preprocessing steps applied. Our pipelines have sufficient overlap in terms of the architecture and the parameters used, hence are all described within this paper. Our code for this work can be found at: https://github.com/ivadomed/ms-challenge-2021.

## 1. Introduction

Magnetic Resonance Imaging (MRI) is routinely used for the diagnosis of multiple sclerosis (MS), which is amongst the most prevalent neurological diseases. There are, however, urgent unmet needs for an earlier diagnosis and better monitoring of MS. In particular, segmenting a new lesion in a patient is extremely helpful for choosing the best treatment for this particular patient early on. Segmenting new lesions requires: (i) MRI data of the same participant across multiple time points and (ii) a radiologist who can identify new lesions by studying the MRI scans across multiple time points. This task is time-consuming and prone to error given the high dimensionality of the data. Moreover, in some cases, multiple contrasts are available, which further augments the difficulty of the visual task. Deep learning (DL) methods have proven useful for the segmentation of MS lesions in the brain [1] and spinal cord [2] MRI scans. However, some challenges remain, including the high class imbalance and out-of-distribution generalization.

In this paper, we propose three pipelines for the segmentation of new lesions from the brain and cervical spinal cord MRI scans. The 3D U-Net [3] architecture is fundamental to each of our three pipelines. Since this architecture was not originally developed for MRI, it required substantial modifications to be compatible with 3D kernels and a segmentation task with high class imbalance.

## 2. Related Work

Recent literature discusses some interesting methods for automatically segmenting MS lesions, thanks to the proliferation of DL techniques and the availability of structured data through open-source challenges. For instance, Kruger et al. [4] present an interesting method for automatically segmenting new MS lesions using a 3D convolutional encoder-decoder architecture. The encoder was pretrained with the task of finding lesions in single timepoint FLAIR scans and these pretrained weights were loaded during the actual training on longitudinal data containing baseline and follow-up FLAIR scans. The network was then tasked with producing a segmentation mask containing the locations of the new lesions. For ensuring stability and convergence during training, deep supervision [5] was introduced by adding segmentation layers in the expansive path of the network.

The winning team in the 2015 challenge on longitudinal white matter lesion segmentation of MS [6] used 3D convolutional neural networks (CNNs) with multi-channel 3D patches of MR volumes were used as the input. A separate CNN was trained for each GT and the final segmentation was obtained by combining the output voxel probabilities of these CNNs.

In most of the DL-based segmentation pipelines, the models are evaluated using binary ground truth (GT) data in the form of 0 and 1 voxel values, which typically fail to capture the rich surrounding information present in the experts' segmentations. This can manifest itself in terms of uncertainty in voxel values [1], inter-expert ambiguity [7], partial volume information (PVE) [8]. As a result, the outputs of these models are uncalibrated and do not generally represent the true confidence of the models' segmentation capability. In this regard, SoftSeg [9] presents the case of using "soft" GT masks instead of the conventional binary ("hard") GTs. An improvement in the Dice score for brain lesion and tumour segmentation tasks was reported when the models were trained with soft GT masks. The potential loss of "soft" (PVE) information by using sharp activation functions (such as sigmoid) as the final layers is also discussed and normalized ReLU is proposed as the better alternative. Therefore, our design choices for the pipelines used in this challenge are primarily inspired by [9].

## 3. Materials and Methods

### 3.1. Data

The dataset for the challenge consists of a total of 100 patients (40 for training, 60 for testing) each with two fluid-attenuated inversion recovery (FLAIR) images (baseline and follow-up), four ground-truth masks each annotated by an expert neuroradiologist, and a consensus ground-truth mask formed via a senior expert neuroradiologist confirming or declining the disputed lesions and fusing the final masks by majority voting. 15 different MRI scanners were used in the acquisition process, and the resulting dataset is highly variable with voxel dimensions ranging from 0.6x0.5x0.5 mm$^3$ to 1.2x1.2x1.2 mm$^3$ (rounded to nearest decimal) and image sizes ranging from (144, 192, 192) to (320, 512, 512) across the two timepoints.

### 3.2. Preprocessing

We use the same preprocessing step described in this section in all of the proposed pipelines in order to increase the quality of the inputs to the pipelines described in the later sections. The preprocessing step is applied independently to each of the 40 patients provided to us in the training set, and it utilizes the following packages with versions provided next to them in parenthesis: FSL (5.0.11), Spinal Cord Toolbox (SCT, 5.3.0), ANTs (2.2.0), and ANIMA (4.0.1). The entire preprocessing stage is described below.

Initially, FLAIR images from each session and their associated GT masks were resampled to the isotropic resolution of 0.5x0.5x0.5 mm$^3$. The target resolution is isotropic because MS lesions have no preferred spatial orientation. The choice of voxel size was a compromise between the segmentation precision and the computational burden. For the next pre-processing steps, we split the analyses between the brain and the spinal cord (SC) given their different shape and co-registration constraints (i.e. quasi-rigid for the brain, non-linear for the spinal cord). We extracted the SC from both sessions and performed an initial co-registration step on the baseline FLAIR image using the segmentation masks of the SC and considering the follow-up FLAIR image as the reference. This first co-registration step was carried out with `sct_register_multimodal` from SCT and it used the slice-by-slice regularized registration algorithm [10] with the mean squares metric and a smoothing kernel of 3x3x3 mm$^3$. The brain was extracted using `bet2` from FSL from the follow-up FLAIR image, and the dilated versions of the brain and SC masks were summed. On observing that new lesions were annotated near the spinal cord region, we decided to incorporate the SC masks also. We then perform a finer co-registration step on the base FLAIR image using the binarized version of the joint brain-SC mask and taking the follow-up FLAIR image as the reference. This second co-registration step was carried out with `antsRegistration` from ANTs, using the neighborhood cross-correlation metric with symmetric normalization transform (SyN), shrink factor of 8x4x2 mm$^3$, and B-Spline interpolation method. The registered baseline and follow-up FLAIR images were then masked using the joint brain-SC mask, and the N4 bias field correction algorithm was applied to both images. The bias field correction step was carried out using `animaN4BiasCorrection` from ANIMA. Finally, the two FLAIR images were cropped using the joint brain-SC mask such that they only include the volume-of-interest.

After performing the above preprocessing step, the maximum image sizes (across all 40 subjects) along the sagittal, coronal, and axial axes in MNI space, were found to be 311, 391, and 487, respectively. In order to assess the performance of the final co-registration and the preprocessing step in general, we visualized the results in a before-after fashion for each of the 40 patients. The visualization is available at this link: https://bit.ly/3nMpXFM.

### 3.3. Data Augmentations

The preprocessed baseline and follow-up FLAIR images with their associated GT masks were initially center cropped according to the maximum image sizes mentioned in the previous subsection, and then divided into subvolumes of size $S$ with a stride-size of $R$. In our pipelines, we have mostly experimented with $S \in \{32, 64, 128, 256\}$ and $R \in \{32, 64, 96, 128\}$. We applied the following data augmentations on the subvolumes while training all of our models: (i) random left-right (lateral) flipping, (ii) random affine transformations from `ivadomed` [11], (iii) random elastic transformation from `ivadomed`, and (iv) random MRI bias-field artifacts from `torchio` [12]. Subsequently, we normalized all images to zero mean and unit variance. Additionally, we use a naive data balancing strategy while training all of our models which duplicates each positive subvolume as many times as required to reach a balanced set (i.e. equivalent numbers) of positive and negative subvolumes. This is applied once before the first epoch in each training process. We observed that balancing the dataset helped with convergence, even though it increased the runtime of the training process. Duplicating subvolumes also poses the potential problem of overfitting, yet the data augmentations described above diversified the samples to a sufficient degree and effectively prevented this problem from occurring.

### 3.4. Challenge pipelines

This section describes the different pipelines we used in this challenge. Each pipeline features an adaptation of the 3D U-Net [3] architecture for the purpose of segmenting MS lesions. A detailed description of each pipeline is provided below.

**Pipeline #1: Modified 3D U-Net**

A U-Net-based architecture was chosen due to its exceptional generalization capabilities achieved by propagating contextual information from the contractive (downsampling) path to the higher resolution layers in the expansive (upsampling) path. The 3D U-Net [3] was proposed as the three-dimensional generalization of U-Net [13] for volumetric segmentation of the *Xenopus* Kidney from sparsely annotated maps. All the 2D operations in the standard U-Net architecture were replaced with their 3D counterparts. We propose to adapt this 3D U-Net model for the challenge by introducing some modifications to the network architecture described below.

The core of the architecture contains four downsampling and upsampling operations. In the contractive path, the blocks at each scale contain 2 sets of 3x3x3 convolutional and instance normalization (IN) layers with strides 2 and 1, respectively, to enable downsampling. A LeakyReLU activation layer is added between each convolutional and IN set. In the expansive path, instead of using transposed convolution layers, the nearest neighbour upsampling is used to avoid the checkerboard artifacts [14]. Each upsampling layer was sandwiched between blocks containing 3x3x3 convolutional of stride 1, IN, and leakyReLU layers. 1x1x1 convolutional layers were used at each scale to match the dimensions of the feature maps concatenated from the contractive path. It is important to note that the original 3D U-Net paper used batch normalization and ReLU activation layers. IN was chosen in this work to account for small batch sizes. The normalized ReLU was used as the final activation layer as inspired by [9], where a ReLU activation is applied to the output and normalized by its maximum value to get an output between 0 and 1, thereby obtaining a full range of prediction values. While the sigmoid layer also squashes its input between 0 and 1, it tends to narrow the range of soft values that carry PVE information required for a good soft prediction. Dice Loss [15] was used as the loss function.

**Pipeline #2: Ensemble of 3D U-Nets**

An ensemble of 4 U-Net models was used in our second pipeline. 3 of these models were Modified 3D U-Net as described in the first pipeline, and the last model was an Attention 3D U-Net as described in more detail in the third pipeline. The final prediction was obtained by taking the soft average of 4 segmentations (one per model). Each U-Net model was trained and validated independently with a different data split, hence diversifying the final prediction in the hopes of achieving better generalization performance.

**Pipeline #3: Attention 3D U-Net with Monte-Carlo Dropout**

Building on top of the first pipeline, our third pipeline features a Modified 3D U-Net supplemented with attention gate (AG) blocks as proposed by [16]. Our motivation to use attention blocks arises from the fact that the challenge dataset contains varying shapes of lesions which constitute a small percentage of the subvolumes they belong to. In this scenario, attention can help the network learn lesion structures more efficiently and in a more informed way by highlighting salient regions and suppressing irrelevant regions.

Typically, using a single set of best weights from the model results in predicted segmentations with high variance. In order to mitigate this issue, we additionally used Monte-Carlo (MC) Dropout [17]. The idea is to use dropout during inference such that when each test input is stochastically forward-passed through the network $T$ times (where $T$ refers to the number of MC samples), we obtain a probability distribution over the output space. By taking the mean of these $T$ samples, we get the averaged segmentation output.

### 3.5. Training & Evaluation

The dataset was split into three parts: a held-out test set of volumes, a validation set of subvolumes, and a training set of subvolumes. First, a random subset of 8 subjects out of 40 was chosen as the held-out test set. Then, the remaining 32 subjects had their subvolumes extracted and a second split was applied where 80% of them were randomly chosen as the training set, and the remaining 20% became the validation set.

In each pipeline, we trained our models with the soft Dice score and saved the model with the best soft Dice score computed on the validation set after 100 epochs. The evaluation of our pipeline, mainly for purposes of model selection, then consisted of two stages: (i) soft and hard Dice score metrics computed on the subvolumes in the validation set, and (ii) Dice score, Jaccard score, positive predictive value (PPV), F1 score, and SurfaceDistance metrics computed on the volumes in the test set as implemented by `animaSegPerfAnalyzer` in ANIMA.